\documentclass[universe,article,submit,moreauthors,pdftex]{mdpi} 

\firstpage{1} 
\pubvolume{1}
\issuenum{1}
\articlenumber{0}
\pubyear{2021}
\copyrightyear{2020}
\datereceived{} 
\dateaccepted{} 
\datepublished{} 
\hreflink{https://doi.org/} 
\usepackage{todonotes}
\usepackage{xspace}
\preto{\abstractkeywords}{\nolinenumbers}

\pdfoutput=1

\newcommand{\agile}{AGILE\xspace}
\newcommand{\fermi}{\textit{Fermi}-LAT\xspace}

\newcommand{\ctools}{\texttt{ctools}\xspace}
\newcommand{\gammapy}{\texttt{Gammapy}\xspace}

\Title{Evolution of Data Formats in Very-High-Energy Gamma-ray Astronomy}

\TitleCitation{Evolution of Data Formats in Very-High-Energy Gamma-ray Astronomy}

\Author{C. Nigro\orcidA{}$^{1}$, T. Hassan\orcidB{}$^{2}$, L. Olivera-Nieto\orcidC{}$^{3}$}

\AuthorNames{Nigro Cosimo}

\AuthorCitation{Nigro Cosimo}

\address{%
$^{1}$ Institut de Física d'Altes Energies (IFAE), The Barcelona Institute of Science and Technology, Campus UAB, 08193 Bellaterra (Barcelona), Spain\\
$^{2}$  Centro de Investigaciones Energéticas, Medioambientales y Tecnológicas (CIEMAT), E-28040 Madrid, Spain\\
$^{3}$  Max Planck Institut f\"{u}r Kernphysik,	Saupfercheckweg 1, 69117 Heidelberg, Germany
}

\corres{Correspondence: cosimo.nigro@ifae.es}

\firstnote{} 
\secondnote{}

\abstract{Most major scientific results produced by ground-based gamma-ray telescopes in the last $30$ years have been obtained by expert members of the collaborations operating these instruments. This is due to the proprietary data and software policies adopted by these collaborations. However, the advent of the next generation of telescopes and their operation as observatories open to the astronomical community, along with a generally increasing demand for open science, confront gamma-ray astronomers with the challenge of sharing their data and analysis tools. As a consequence, in the last few years, the development of open-source science tools has progressed in parallel with the endeavour to define a standardised data format for astronomical gamma-ray data. The latter constitutes the main topic of this review. Common data specifications provide equally important benefits to the current and future generation of gamma-ray instruments: they allow the data from different instruments, including legacy data from decommissioned telescopes, to be easily combined and analysed within the same software framework. In addition, standardised data accessible to the public, and analysable with open-source software, grant fully-reproducible results. In this article we provide an overview of the evolution of the data format for gamma-ray astronomical data, focusing on its progression from private and diverse specifications to prototypical open and standardised ones. The latter have already been successfully employed in a number of publications paving the way to the analysis of data from the next generation of gamma-ray instruments, and to an open and reproducible way of conducting gamma-ray astronomy.}


\begin{document}
\section{Introduction}
Gamma-ray astronomy, currently observing the non-thermal universe over more than 7 decades in energy, is conducted with different classes of instruments operating in two complementary energy ranges \cite{funk_2015}. Space-borne telescopes, sensitive in the so-called high-energy regime (HE, $100\,{\rm MeV} < E < 100\,{\rm GeV}$), directly detect the gamma rays through their pair-conversion in an instrumented volume \cite{thompson_2015}. Ground-based telescopes, sensitive in the so-called very-high-energy regime (VHE, $E > 100\,{\rm GeV}$), detect the particle cascade (or shower) generated by gamma rays interacting with atmospheric nuclei (via ${\rm e}^{\pm}$ pair production and Bremsstrahlung) using two different techniques \cite{de_narouis_2015}. Imaging Atmospheric Cherenkov Telescopes (IACTs) use a large reflector ($\sim 10\,{\rm m}$) and a photomultiplier camera to image the Cherenkov light emitted by the charged component of the shower. Particle samplers rely on an array of detectors (distributed over a surface up to a $\sim {\rm km}^2$) to directly sample the charged component using, for example, scintillators or water tanks in which further Cherenkov light is produced and detected (water cherenkov detectors, WCD). VHE astroparticle physics will be revolutionised in this decade by an upcoming generation of ground-based instruments built with the objective to improve by an order of magnitude the sensitivity of the current ones: the Cherenkov Telescope Array (CTA) \cite{science_cta} for IACTs; the Large High Altitude Air Shower Observatory (LHAASO) \cite{science_lhaaso} and the Southern Wide-field Gamma-ray Observatory (SWGO) \cite{science_swgo} for particle samplers.
\par
Beside different detection techniques, the current generation of HE and VHE instruments adopt distinct data and software policies. As typical for space observatories, HE gamma-ray telescopes retained their data proprietary for a limited amount of time (usually one year) before releasing them publicly. This has been the case for both currently operating HE gamma-ray telescopes: the \textit{Fermi} Large Area Telescope (\fermi)\cite{fermi_data_policy} and the Astrorivelatore Gamma ad Immagini Leggero (\agile) \cite{agile_data_center}. Their data are nowadays made promptly available via web-based platforms, referred to as science data centers, providing astronomers with an interface to retrieve the data and the science tools to perform their analysis \cite{fermi_ssc, agile_data_center_web}. VHE telescopes of the current generation, on the other hand, have been operated under more strict data and software policies. Telescopes like the High Energy Stereoscopic System (H.E.S.S.) or the Major Atmospheric Gamma-ray Imaging Cherenkov (MAGIC) have traditionally produced scientific results with proprietary data and closed-source software \cite{mars, hap-fr}. Few examples of public VHE data or software exist, worth mentioning are: the Very Energetic Radiation Imaging Telescope Array System (VERITAS), that has publicly released under an open-source license one of its analysis chains \cite{eventdisplay}; the First g-Apd Cherenkov Telescope (FACT), that has made public its analysis chain \cite{fact_tools}, a small sub-sample of its data and quick-look analyses results on all the data collected \cite{fact_open_data}; and the High-Altitude Water Cherenkov (HAWC), that has provided some high-level data, mostly meant to reproduce results of major publications \cite{hawc_public_datasets}. More recent efforts of data sharing in a standardised format will be covered later in this review. Generally speaking, beside sparse endeavours, VHE gamma-ray data largely remain inaccessible to astronomers outside the collaborations gathering them. This situation is bound to change with the forthcoming CTA that will represent the first gamma-ray experiment operated as a proposal-driven open observatory \cite{cta_data_management}. External scientists will be able to submit observational proposals; data will be proprietary to the principal investigators typically for one year and then released to the public. This implies, as in the case of HE gamma-ray instruments, the necessity to produce accessible data and tools for users external to the collaboration to perform their scientific analyses.
\par 
In light of these requirements, VHE gamma-ray astronomers have started developing open-source data-analysis tools (e.g. \texttt{ctools} \cite{ctools} and \texttt{Gammapy} \cite{gammapy}) and, in parallel, a standardised format for astronomical gamma-ray data. This review will focus on the latter. The data level expected to be shared by the next generation of VHE observatories with external observers (as already routinely done by \fermi and \agile) is a \textit{high} data level whose purpose is the production of scientific results (i.e. measurement of the properties of an astrophysical source: flux, morphology, etc.). It contains a reduced amount of information compared to the \textit{low} (or calibrated) data level strictly connected with the particular detection or analysis technique. Specifically it contains lists of detected photons with their estimated physical observables (energy, direction, etc.) and a characterisation of the response of the system. It is abstract enough to represent data from instruments employing diverse detection techniques such as IACT and WCD. Being difficult to detach the discussion on high-level data format from the software provided to analyse it, we might comment as well upon aspects of software development and policies. 
\par
This review is thus structured: the progression of the data format from previous specifications is discussed in Sect.~\ref{sec:gadf}, along with its current status and working principles. In Sect.~\ref{sec:applications} we review some projects that have already successfully employed the format, either to validate the capabilities of the science tools, to illustrate the possibility of multi-instrument analysis with current gamma-ray instruments and to extend the format to particle samplers. In Sect.~\ref{sec:future} we gather some ideas for the future of the format and its possible expansion. We provide our conclusions in Sect.~\ref{sec:conclusions}.

\section{Data formats for very-high-energy gamma-ray astronomy}
\label{sec:gadf}

\begin{figure*}
\includegraphics[scale=1.0]{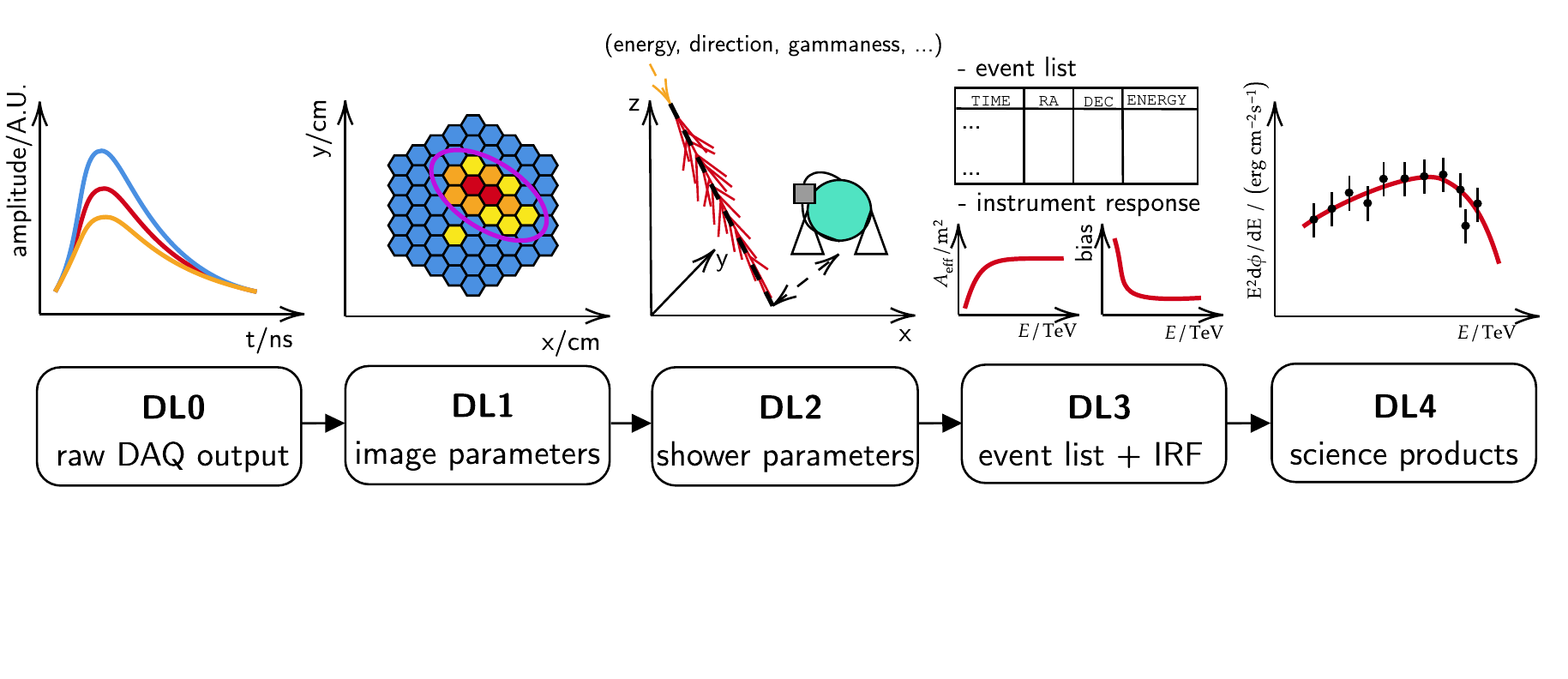}
\caption{Schematisation of the progressive data reduction and data levels of an IACT. Raw data contain the signal sampled from the photomultipliers at the occurrence of a trigger event (Data Level 0). Calibrated data (Data Level 1) contain the pixelated image of the Cherenkov light of the shower. The latter can be parametrised with few geometrical quantities and used to determine the observables of the original shower, including its probability of being a gamma-ray shower (Data Level 2). The detected events can be gathered in a list of gamma ray candidates, together with the functions representing the response of the system (the so-called instrument response function, IRF) e.g. the collection area of the system as a function of the energy or the bias of its energy reconstruction (Data Level 3). This information can be used to perform a statistical analysis obtaining the so-called science products, in this case the spectrum of the source (Data Level 4).}
\label{fig:data_levels}
\end{figure*}

\subsection{Background: data model in the current generation of VHE instruments}
\label{sec:data_levels_intro}
VHE gamma-ray astronomy inherited, along with the hardware techniques, the software solutions of particle physics. In the late 1990s and early 2000s, \texttt{C++} and the \texttt{ROOT} \cite{root} framework dominated the field. Hence, software for VHE data reduction and analysis has been mostly built in this environment. As already commented, even if some of these tools are accessible, little documentation is publicly available about the private analysis chains and the data they produce. Nonetheless, from the available material, a common data reduction workflow can be inferred for VHE gamma-ray telescope, sketched in Fig.~\ref{fig:data_levels}. 
\par
In the case of an IACT, the raw output of the data acquisition typically consists of binary files containing the waveforms of all the camera pixels, sampled at the occurrence of a trigger event. The raw data are reduced to a list of quantities per pixel (e.g. charge and arrival time) aggregated in the so-called \textit{calibrated} files with size of several ${\rm GB}$ for each observational run, typically around $\sim 30\,{\min}$ (in what follows the sizes indicated per each data level are taken from \cite{vegas}, so they refer to VERITAS. One can compare with similar figures reported in \cite{nigro_phd} for MAGIC). The Cherenkov light of the shower typically illuminates few pixels in the camera, this pixelated image, representing the distribution of Cherenkov photons, can be parametrised with simple geometrical quantities \cite{hillas} connected to the shower properties. Image parameters can be fed, at the next data level, to algorithms estimating these properties (e.g. energy and direction of the primary) and classifying the showers initiated by gamma rays against those initiated by cosmic rays, the irreducible background of ground-based gamma-ray telescopes. In the case of particle samplers such as WCD, the data reduction workflow is similar but instead of camera images, the information is extracted from the pattern in the charge deposited by the shower across the array, as well as from its time evolution. Raw parameters derived from this charge distribution are fed into reconstruction algorithms that in turn, estimate the relevant shower parameters, like those mentioned above (see~\cite{hawc-crab-2017} for an overview of the HAWC data reduction pipeline). Having estimated the properties of the shower and of the primary particle generating it, a list of gamma-ray candidates can hence be assembled at the next data level. 
\par
At this stage, the information stored within the data products, generally denoted as \textit{high-level}, is independent of the detection technique as well as the calibration and analysis methods. High-level data typically consist of a list of gamma-ray events along with a parametrisation of the response of the system, the so-called instrument response function (IRF). The latter provides the information necessary to perform a statistical analysis estimating, for example, the significance of the signal, the flux spectrum or the light curve of the source, which we refer to as science products. 
\par
All along the current-generation closed-source analysis chains the data, progressively reduced, are stored in the format associated with the \texttt{ROOT} framework, with each collaboration reiterating the effort of defining custom specifications for a data model that shares several commonalities between different experiments. Moreover, even if readable via \texttt{ROOT}, the content of these data products cannot be interpreted by a non-expert analyser. There are noticeable efforts to provide analysis tools wrapping these diverse analysis software like the Multi-Mission Maximum Likelihood framework \cite{3ML}. The ultimate limitation of these tools is though the availability of the experiments to expose their closed-source software and data format and the necessity to implement a new plug-in for each of the instruments considered. 
\par
Without a common data model or a general software tool oriented to external users, the current generation of VHE instruments faces different concerns in different time perspectives. At present, multi-instrument analyses simply cannot be performed within a common analysis framework using their proprietary data products. For what concerns the future, as the end of their operation approaches, it is worth to start considering the access to the wealth of data they gathered. If their \textit{legacy} data are to be made public then a release in their original format will make necessary a release of the analysis software as well, which in turn has to be maintained. Beside not being designed for the usage by a large community, this software can rely on libraries that will eventually become deprecated.

\subsection{GADF: A unifying effort}

\begin{figure}
    \centering
    \includegraphics[scale=0.15]{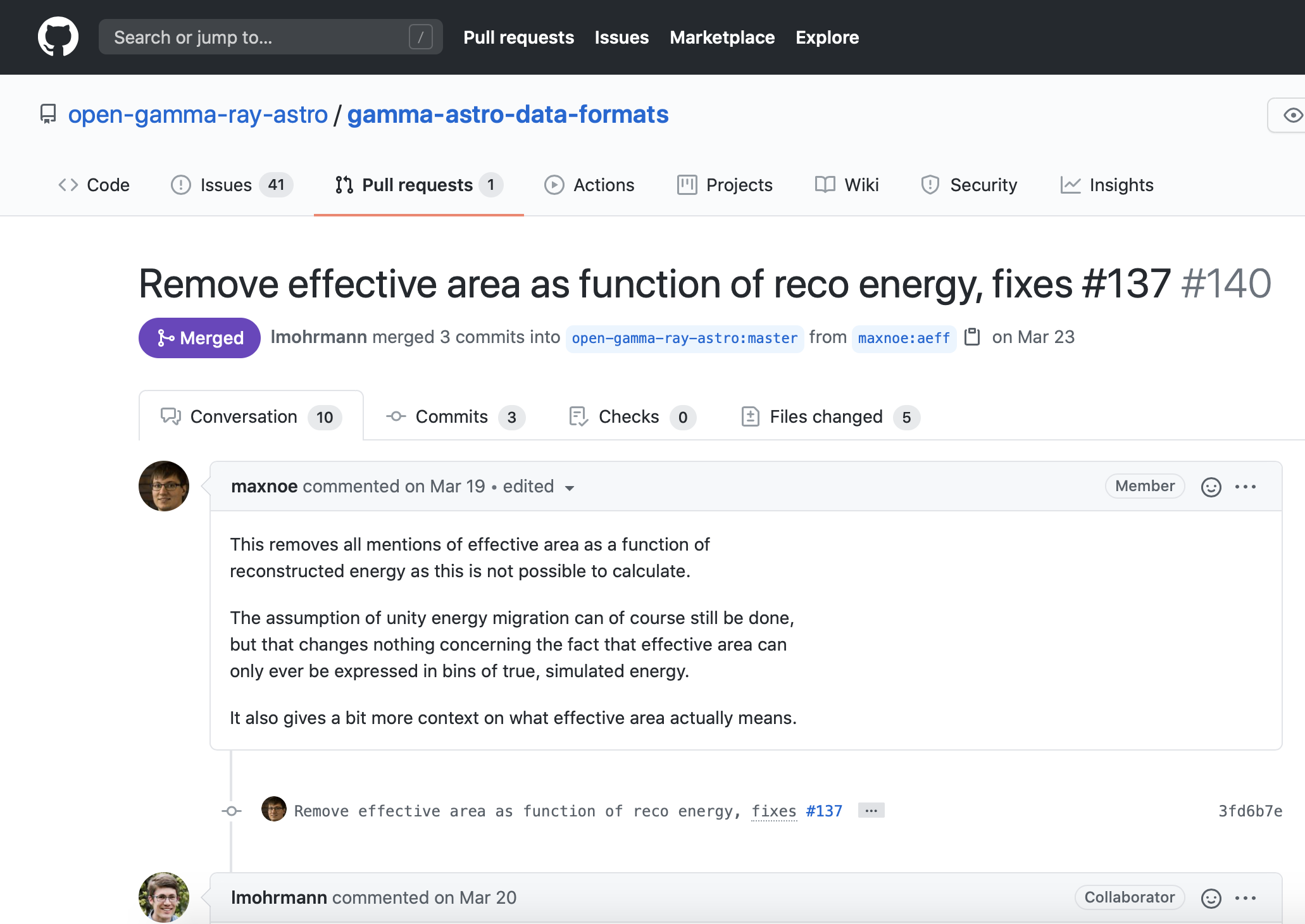}
    \includegraphics[scale=0.15]{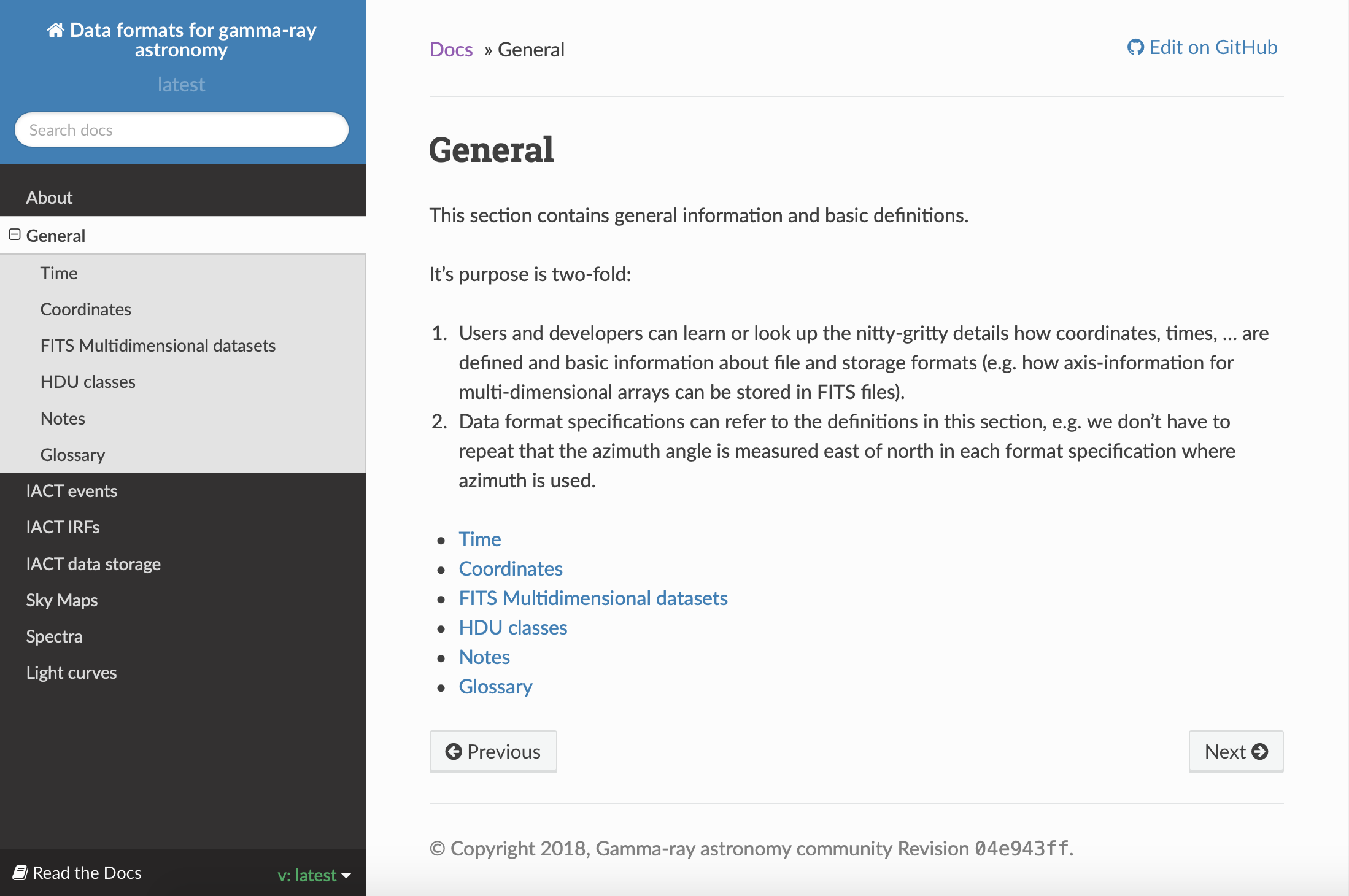}
    \caption{\textit{Left}: \texttt{GitHub} repository hosting the development of the \textit{Data Formats for Gamma-ray Astronomy} specifications. \textit{Right}: The repository contains a documentation written in \texttt{sphinx} whose \texttt{html} version can be explored on \texttt{readthedocs}.}
    \label{fig:gadf}
\end{figure}
In the second half of the 2010s, partly to prototype the high-level data format of the forthcoming CTA and partly to exploit the newly available open-source data-analysis software like \ctools and \gammapy, VHE astronomers started to explore several software-independent implementations of these high-level data. In 2016, in order to coordinate the parallel efforts and to foster the definition of a common and standardised data model, the \textit{Data Formats for Gamma-ray Astronomy} forum (shortly referred to as the ``gamma astro data formats'', GADF) \cite{gadf} was established. A community-driven initiative, the GADF consists of a documentation \cite{gadf_docs} hosted on \texttt{GitHub} \cite{gadf_github} (Fig~\ref{fig:gadf}), specifying the naming scheme, the content, and the metadata of the files containing high-level gamma-ray observations. Though high-level products are the focus of the initiative, specifications for science products are also under discussion. The documentation, openly provided with a Creative Commons Attribution 4.0 license, evolves with the typical \texttt{GitHub} workflow: any interested user can propose changes via \textit{issues} that will be discussed among the active members of the initiative, and implemented via \textit{pull requests} that will be ultimately merged once a consensus is reached. Despite the bias towards IACTs, the flexible development of the format allows to accommodate data from other types of instruments, such as space-borne telescopes or WCD. The format has achieved a stable definition and counts already two minor releases, the present being \texttt{0.2} \cite{gadf_zenodo}. 

\subsubsection{Format specifications}
\begin{figure}
    \centering
    \includegraphics[scale=0.4]{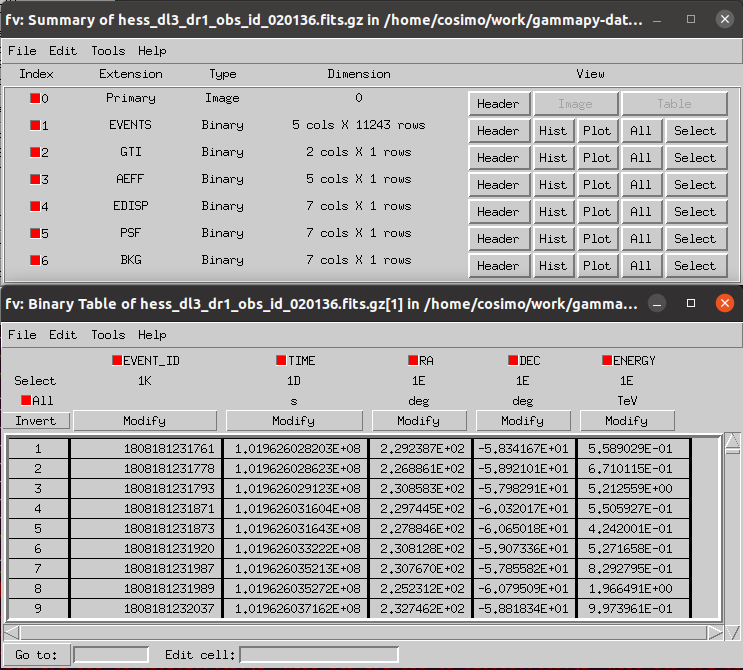}
    \caption{Example of a DL3 file compliant with the GADF specifications. Top: the header data units (or extensions) of the file contain the event lists, under (\texttt{EVENTS}), followed by those representing the good time intervals (\texttt{GTI}) and the instrument response components: effective area (\texttt{AEFF}), energy dispersion (\texttt{EDISP}), point spread function (\texttt{PSF}) and background (\texttt{BKG}). Bottom: event list table and its content.}
    \label{fig:dl3_file}
\end{figure}
This section illustrates the guiding principles adopted in the development of the GADF specifications, gives an overview of their actual content and highlights the features that make them generalisable to different gamma-ray instruments. The first version of the GADF was designed for IACT, since the major contributors were VHE astronomers preparing for CTA. The data model and the breakdown of the data levels foreseen for CTA are presented in \cite{data_model_issue}, introducing the following naming convention (see also Fig.~\ref{fig:data_levels}): the raw output of the data acquisition is defined as data level 0 (DL0); calibrated files as data level 1 (DL1); reconstructed shower parameters as data level 2 (DL2); sets of selected gamma-ray events and the instrument response as data level 3 (DL3); science products (spectra, light curves, sky maps) as data level 4 (DL4), and observatory results as catalogues such as data level 5 (DL5). This nomenclature is used within the GADF and will be also adopted in the following text.
\par
As the GADF is currently the only provider of standardised specifications for high-level VHE gamma-ray data, science tools as \ctools and \gammapy base their data structures on them. Compatibility with open-source data-analysis software is not the only objective of the standardisation effort. One of the guiding principles of the GADF is to produce data whose content is clearly documented and easy to interpret. The file format chosen to host the data is the Flexible Image Transport System (FITS) \cite{fits}, representing a long-time standard in astronomy at all wavelengths. Another fundamental requirement in the design of the data specifications was to rely as much as possible on already well-established standards used in other FITS files productions, such as those by the missions gathered under NASA's High Energy Astrophysics Science Archive Research Center (HEASARC) \cite{heasarc}. NASA's Office of Guest Investigator Program (OGIP) FITS working group \cite{ogip} already disseminates to the high-energy astrophysics community recommendations on FITS data productions. These include standards on keyword usage in metadata, on storage of time information, representation of response functions that the GADF extensively follows. The adherence of the GADF to widely used standards ensures additional compatibility with tools already in use by the high-energy astrophysicists like the \texttt{FTOOLS} \cite{ftools}.
\par
As pointed out, the aim of the GADF initiative was to produce specifications for high-level data, therefore, it mostly focuses on the DL3. Nonetheless, the forum discusses data levels higher than the DL3. For example, the OGIP spectral file format \cite{ogip_spectral_format} is adopted to represent VHE gamma-ray one-dimensional (energy-dependent) spectral data. The compatibility with the OGIP standards ensures that DL3 products can be reduced to spectral data digestible by other established multi-mission analysis tools such as \texttt{sherpa} \cite{sherpa_1, sherpa_2}. Prototypical specifications for DL4 (such as sky maps, flux points and lightcurves) are under discussion and not yet stable.

\subsubsection{GADF DL3 data}
\begin{figure}
    \centering
    \includegraphics[scale=0.35]{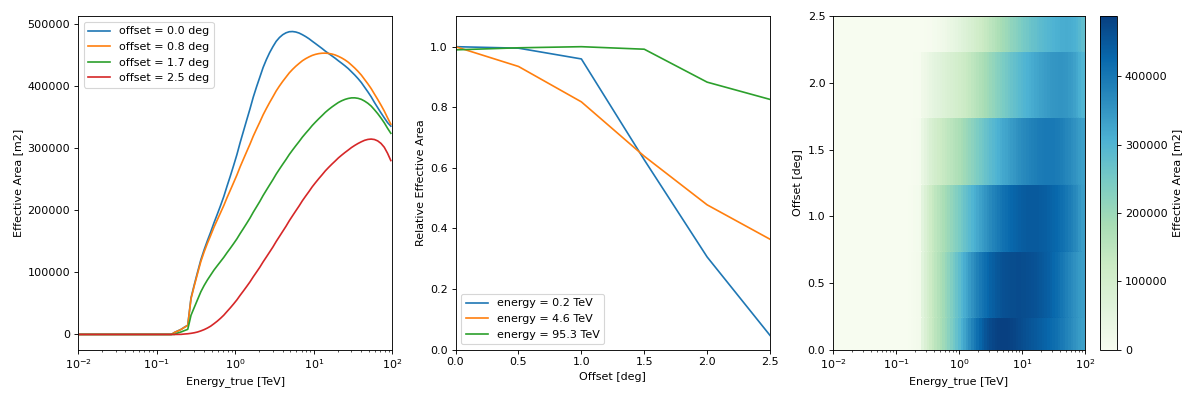}
    \caption{Example of visualisation of the effective area of a IACT and its dependency on energy and offset angle. The IRF component is read from DL3 files and displayed using \gammapy.}
    \label{fig:aeff}
\end{figure}
The DL3 is the data level that contains a list of gamma-ray event candidates and the response of the system. All the information in the DL3 files is therefore post-calibration, i.e. already incorporating all the low-level information related to the detector (calibration, gain corrections, digital-count-to-photo-electron conversion) that is hence omitted. A FITS file consists of many extensions, called header data units (HDUs). Each HDU is composed by a header unit, typically containing metadata, and a data unit, containing a n-dimensional array (an image) or a table (in ASCII or binary format). All data units in DL3 files are stored as binary tables. 
\par
One of the file extensions contains the event list and, in the associated data unit, a flat table with a column for each event property (see Fig~\ref{fig:dl3_file}). In the current specifications columns listing the events identification number (in the DAQ system), energy, sky coordinates (right ascension and declination) and timestamp are mandatory. Optional columns might include results of the classification algorithms (e.g. a \textit{gammanness} score) and quantities related to the reconstruction (e.g. image or shower parameters). Each file corresponds to a single observing run, therefore the events header unit contains the identification number of the data acquisition run, the type and number of telescopes used in the observation, information about the location of the instrument and its observation mode along with time and duration of the observation. Another HDU is dedicated to a list of good time intervals (GTI), specifying the time periods within the event lists with adequate scientific quality.
\par
The response of the system is needed to properly relate the reconstructed events with astrophysical source properties. It is assumed that this response can be factorised in different components. The components considered are: the effective area, describing the acceptance of the system to gamma-ray events; the energy dispersion (or migration matrix), describing the probability distribution of the energy estimator and the point spread function (PSF), describing the probability distribution of the direction estimator. The background rate (measuring the rate of cosmic ray events misclassified as gamma rays) might be included among the IRF components, however it is not mandatory. The IRF components depend on observational (e.g. atmospheric conditions, zenith and azimuth angle of the pointing) and physical quantities (e.g. the energy or direction of the showers). The IRF components considered in the format are valid for a single exposure, which is tipically defined by constant observational conditions (e.g. zenith range, atmospheric quality, etc.), hence considering any such dependency of the IRF averaged out. In the current specifications, the dependencies on physical quantities considered are the photon energy and the offset of its position from the centre of the instrument field of view (a response symmetric with the offset coordinate is assumed). As an example, Fig.~\ref{fig:aeff} illustrates the energy and offset dependency of the effective area component for a H.E.S.S. observation stored in the GADF DL3 format. IRF components are not stored in flat tables: energy and offset bin edges are stored in separate columns, and a last column contains a multi-dimensional array corresponding to the response in each bin. OGIP specifications are followed in storing both events and IRF components.

\section{Projects successfully using the standardised data format}
\label{sec:applications}
To illustrate the maturity of the GADF standardisation effort, we review, in the following sections, projects that have successfully employed its specifications. 

\subsection{The H.E.S.S. first public test data release}
\label{sec:hess_dl3_dr1}

\begin{specialtable}
\caption{Content of the H.E.S.S. Data Level 3 Data Release 1.}
\label{tab:hess_dl3_dr1_sources}
\begin{tabular}{|l|l|l|l|}
\toprule
source & type & case study & time / h \\
\midrule
Crab Nebula & Pulsar Wind Nebula & point-like, steady source & 1.9 \\ 
PKS~2155-304 & Blazar & point-like, variable source & 9.8 \\
MSH~15-52 & Pulsar Wind Nebula & small-extension, steady source & 9.1 \\
RX~J1713.7-3946 & Supernova Remnant & large-extension, steady source & 7.0 \\
off data & various & background estimation & 20.7 \\
\bottomrule
\end{tabular}
\end{specialtable}

The High Energy Stereoscopic System (H.E.S.S.) collaboration was the first to publicly release a test dataset in a DL3 format compliant with the GADF specifications. Few observations, amounting roughly to $\sim 50 {\rm h}$ of observation time, gathered between 2004 and 2008 were published in the so called H.E.S.S. DL3 Data Release 1 (H.E.S.S. DL3 DR1) \cite{hess_dl3_dr1_arxiv, hess_dl3_dr1_web} to promote the standardisation effort but also to allow to test the open-source science tools in development with actual IACT data. The data release contains $30\,{\rm h}$ of observations of sources representing different galactic and extragalactic science cases, and $20\,{\rm h}$ of observations of field of views empty of known gamma-ray emitters, also labelled as \textit{off} data, to be used for background estimation. Table~\ref{tab:hess_dl3_dr1_sources} summarises the content of this data release.

\subsection{The joint-crab project}
\label{sec:joint_crab}
\begin{figure}
    \centering
    \includegraphics[width=0.35\textwidth]{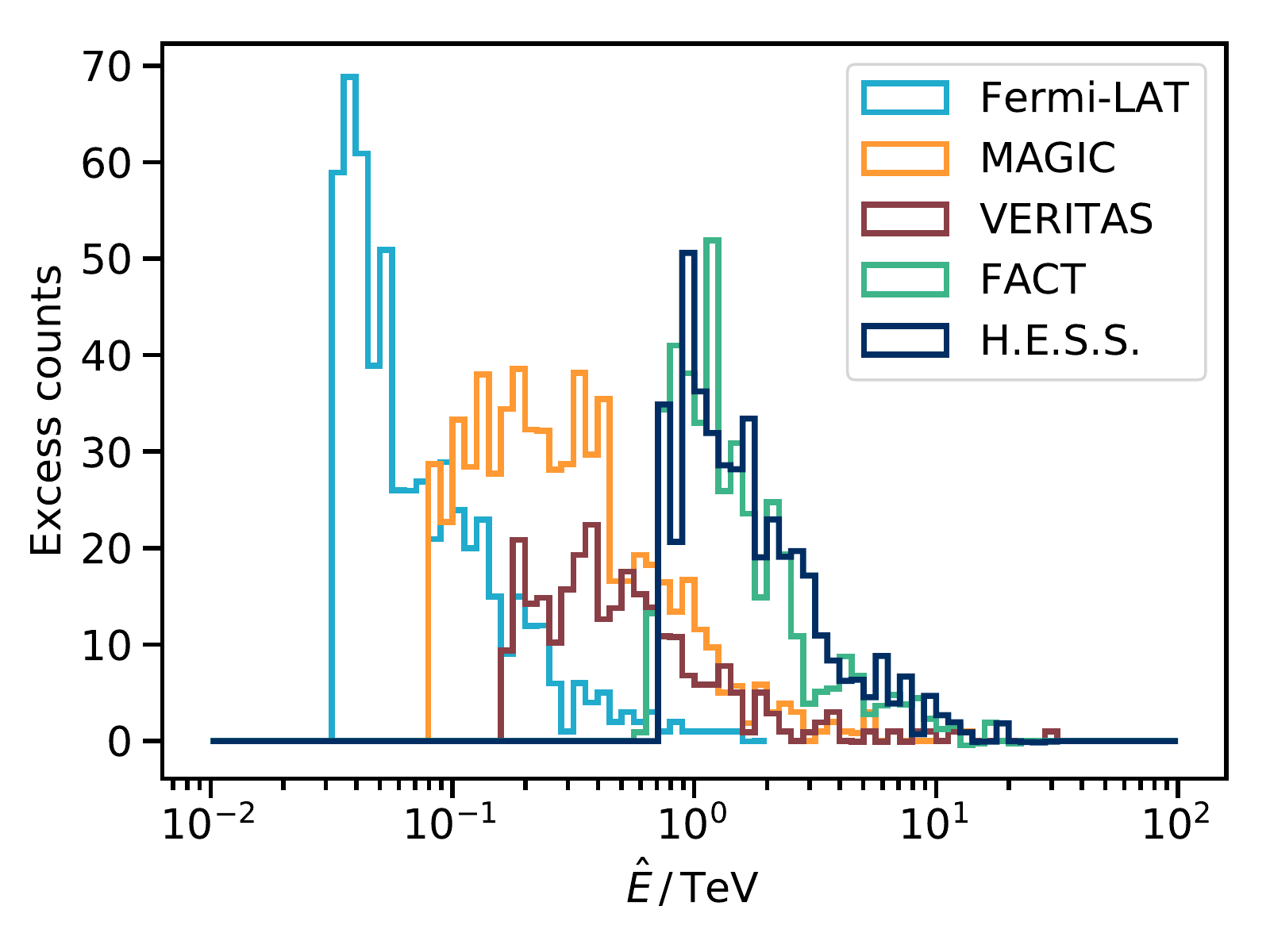}
    \includegraphics[width=0.35\textwidth]{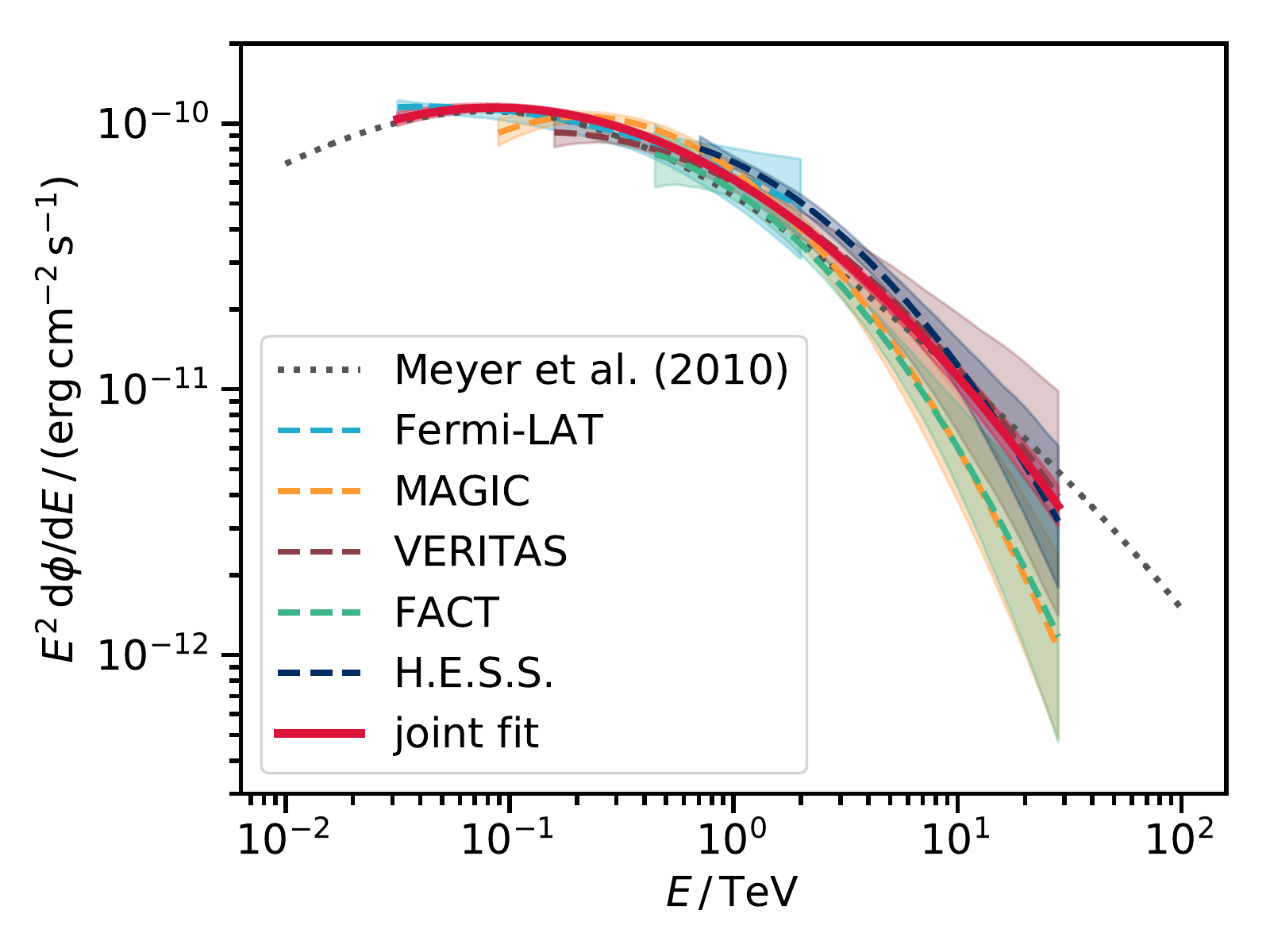}
    \caption{\textit{Left}: Source counts vs estimated energy extracted via aperture photometry, per each of the instrument datasets in \cite{joint_crab}. \textit{Right}: Estimated flux spectrum of the Crab Nebula obtained from the individual instrument datasets (same color code as in the figure on the left) and considering all the datasets in the same joint likelihood (red). The grey dashed line represents a bibliographic reference. In all cases the analytical flux model considered in the likelihood is a curved power-law. Figures from \cite{joint_crab}.}
    \label{fig:crab_sed}
\end{figure}
With multi-instrument analyses being one of the main objectives of the standardisation effort, after the first public release of GADF-compliant DL3 data, the next step in the format validation would have naturally been the combination of data from different experiments. In the so-called \textit{joint-crab} project \cite{joint_crab}, Crab Nebula observations from \textit{Fermi}-LAT and four of the currently operating IACTs, produced in a GADF-compliant format, were combined in the first multi-instrument and fully-reproducible gamma-ray analysis. The datasets used were:
\begin{itemize}
    \item $7\,{\rm yr}$ of \textit{Fermi}-LAT observations, obtained in the custom high-level, DL3, format with which they are publicly released. They were reduced, before the final statistical analysis, to OGIP spectral data;
    \item $2\,{\rm h}$ of H.E.S.S. observations selected from the H.E.S.S. DL3 DR1 (see Sect.~\ref{sec:hess_dl3_dr1});
    \item $40\,{\rm min}$ of MAGIC observations produced and released specifically for this project;
    \item $40\,{\rm min}$ of VERITAS observations produced and released specifically for this project;
    \item $10\,{\rm h}$ of FACT observations from their already public data sample (see Sect.~\ref{sec:data_levels_intro}).
\end{itemize}
To illustrate a prototypical analysis example, the Crab Nebula spectrum (Fig.~\ref{fig:crab_sed} right) was estimated combining all the observations in an energy-dependent (or one-dimensional) joint binned likelihood. In this analysis technique, classically employed by IACT, source and background events are extracted via aperture photometry (Fig.~\ref{fig:crab_sed} left) and then an energy-dependent analytical flux model is folded with the response of the system to estimate the number of counts maximising the Poissonian likelihood describing the counts in each energy bin. The \textit{joint-crab} project relied only on open-source software for its statistical analyses (\texttt{Gammapy}). Datasets, scripts reproducing all the analysis steps and tutorial notebooks are publicly provided on \texttt{GitHub} \cite{joint_crab_github}, along with a \texttt{conda} environment freezing the exact dependencies used in the paper and a docker container \cite{joint_crab_docker} to guarantee a long-term reproducibility. The entire package was also archived on \texttt{zenodo} \cite{joint_crab_zenodo}. Given the approach proposed and the assets openly made available, this work not only implements the first fully-reproducible gamma-ray analysis but also constitutes the first joint public release of IACT DL3 data. 

\subsection{Analysis of the H.E.S.S. public data release with ctools}
\label{sec:ctools_hess_dl3}
\label{sec:validation}
\begin{figure}
    \centering
    \includegraphics[width=0.39\textwidth]{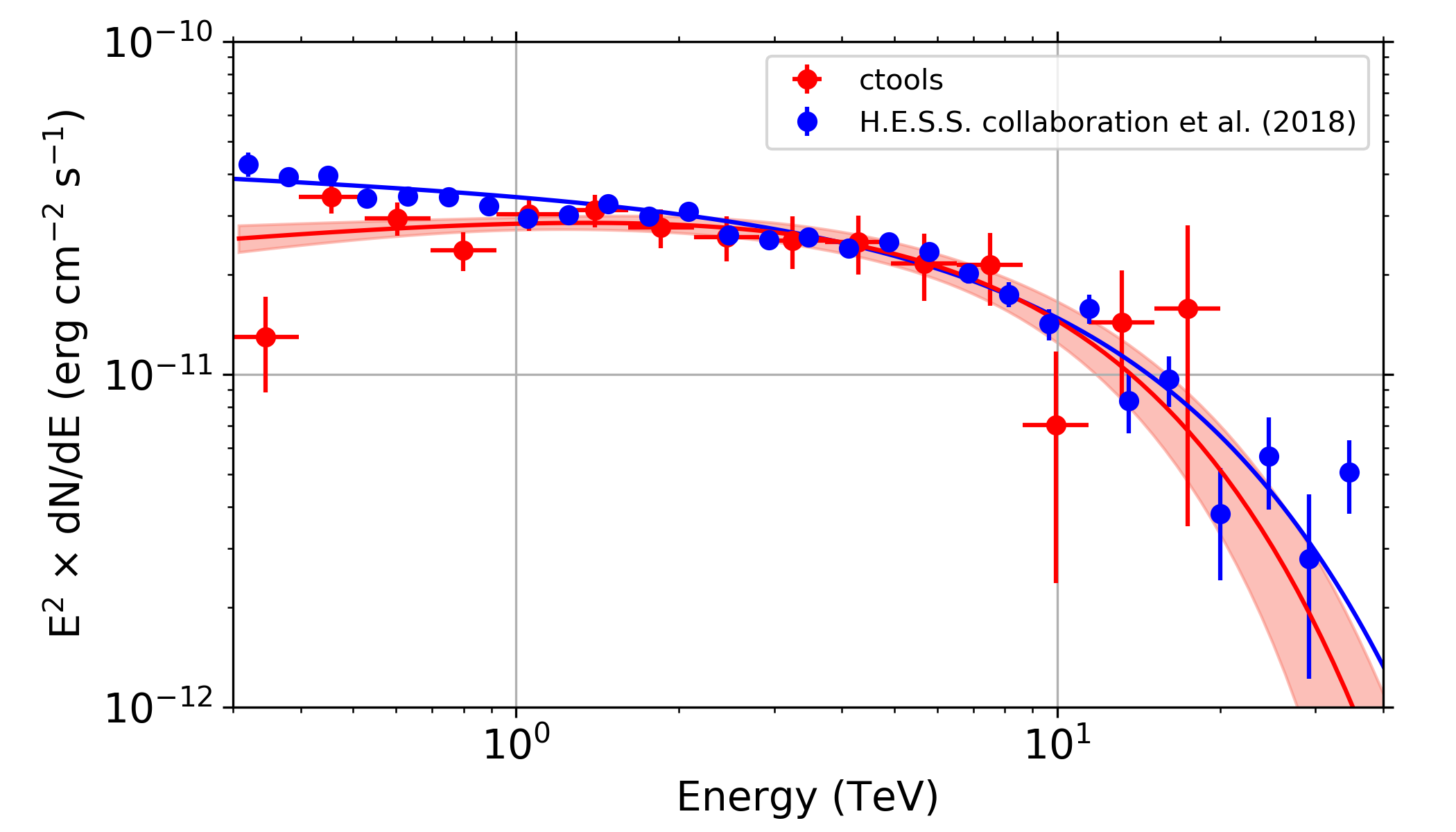}
    \includegraphics[width=0.33\textwidth]{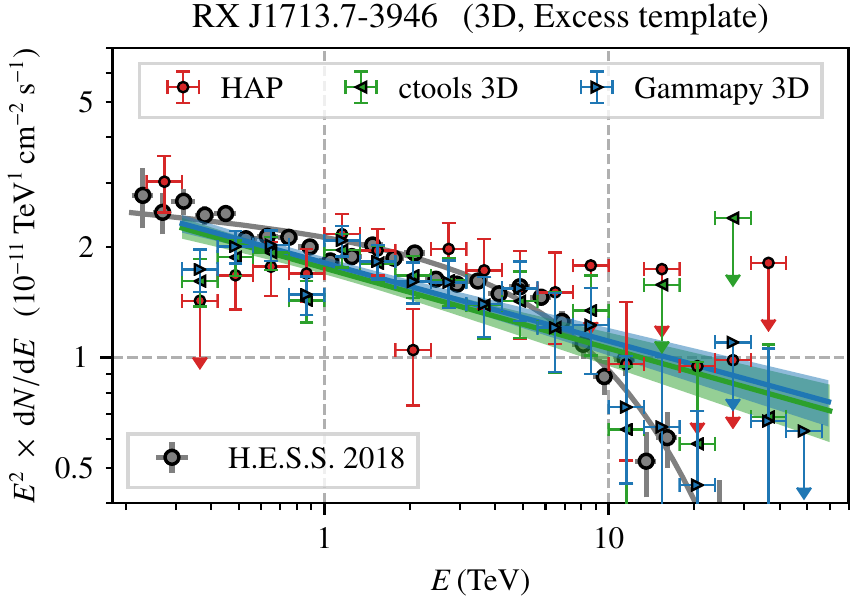}
    \caption{Flux spectrum of the extended gamma-ray source RXJ1713.7-3946. \textit{Left}: Comparison of the result obtained from the H.E.S.S. DL3 DR1 using \texttt{ctools}' three-dimensional unbinned likelihood analysis (red) against the literature (blue). Figure from \cite{ctools_hess_dl3}. \textit{Right}: Comparison of the result obtained from the H.E.S.S. DL3 DR1 using \texttt{ctools}' (green) and \texttt{Gammapy}'s (blue) three-dimensional likelihood analyses against a result obtained on the very same data sample with the H.E.S.S. private analysis chain (red), performing a one-dimensional analysis. A bibliographic reference, the same as in the figure on the left, is given in grey. Figure from \cite{validation_science_tools}.}
    \label{fig:rxj_science_tools}
\end{figure}
Besides evolving in parallel with the GADF, the open-source science tools can recognise data with its specifications as input. In \cite{ctools_hess_dl3}, the H.E.S.S. DL3 DR1 (Sect.~\ref{sec:hess_dl3_dr1}) was used to test the capabilities of \ctools, until then mainly used to analyse simulated CTA observations and calculate prospects for its observational capabilities. The authors presented a method to build a parametric model describing the spatial and spectral distribution of the background events in the H.E.S.S. DL3 DR1. The latter was used to perform a spectro-morphological (three-dimensional) analysis estimating the spectrum of the 4 sources included in the data release. Differently than in the one-dimensional analysis described in Sect.~\ref{sec:joint_crab}, the sources positions and morphology are included among the parameters of the model used to estimate the flux. Source and background counts are not separated, rather the background is included among the components of a model that in this case predicts the flux in the entire field of view, allowing to take into account multiple sources at a time (see \cite{ctools} Sect.~2 for a detailed explanation). This approach has been successfully used by the \fermi collaboration for all its scientific publications. The results of binned and unbinned three-dimensional likelihood analyses are compared against the simpler one-dimensional binned analysis, also implemented in \ctools, and against bibliographic references obtained from the same sources. The consistency of the results obtained with \ctools with the different statistical methods applied and with the literature (see Fig.~\ref{fig:rxj_science_tools} left) testifies the maturity not only of the science tool, but also of the GADF scheme that correctly encapsulates all the information needed for correct reproduction of scientific results. The paper finally illustrates the capability of \ctools, being built on the \texttt{gammalib} library \cite{ctools}, to simultaneously analyse gamma-ray data with different specifications, i.e. to analyse \fermi data in their own high-level format (without the reduction described in Sect.~\ref{sec:joint_crab}) and IACT DL3 data compliant with the GADF specifications.

\subsection{Validation of open-source science tools and background model construction in $\gamma$-ray astronomy}
Expanding on the project described in Sect.~\ref{sec:ctools_hess_dl3}, \cite{validation_science_tools} aims at testing both \texttt{Gammapy} and \texttt{ctools} using the H.E.S.S. DL3 DR1. The results of the one-dimensional and three-dimensional analyses provided by both science tools are validated against each other. For the three-dimensional analysis, a novel background model is used, not parameterised from the \textit{off} sources within the H.E.S.S. DL3 DR1, but built using $\sim 4000\,{\rm h}$ hours of H.E.S.S. private observations. For this work the results of the science tools are validated not only against the literature, but also against the results obtained with one of the closed-source analysis chains of the H.E.S.S. collaboration, performing a classical one-dimensional analysis on the exact same observations included in the H.E.S.S. data release (see Fig.~\ref{fig:rxj_science_tools} right). The agreement of the results of the different science tools among them and with the private analysis chain represents a landmark in the analysis tools and data formats validation for future VHE gamma-ray analyses. 

\subsection{Open and standardized formats for $\gamma$-ray analysis applied to HAWC observatory data}
\label{subsec:hawc}
\begin{figure}
    \centering
    \includegraphics[scale=0.45]{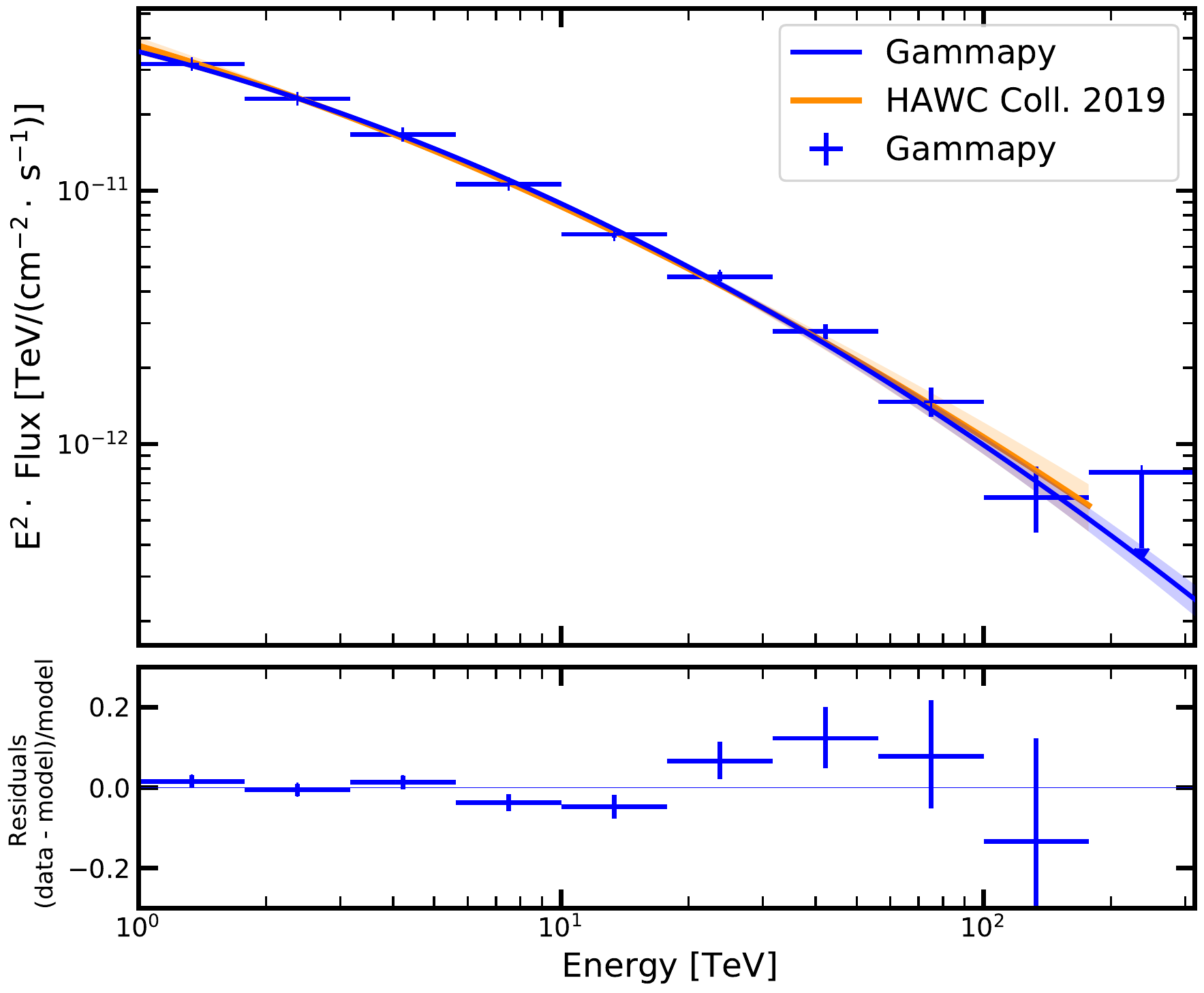}
    \caption{Estimated flux spectrum of the Crab Nebula derived from the GADF data production using \texttt{Gammapy} (blue) compared against the reference HAWC spectrum from~\cite{hawc-crab} (orange). The bottom pannel shows the residual comparison of the obtained flux points with the reference spectrum. Figure from~\cite{gadf_hawc}.}
    \label{fig:crab_hawc}
\end{figure}

The GADF specifications were primarily developed by and for the IACT community. However, due to their generality, it is possible to use them to format data from WCD, such as the HAWC observatory, as shown by \cite{gadf_hawc}. In this work the authors presented the first GADF-compliant production of event lists and instrument response functions for a ground-based wide-field instrument. These data products were then used to reproduce with excellent agreement the published spectrum of the Crab Nebula as measured by HAWC. This result, shown in Fig.~\ref{fig:crab_hawc}, was obtained using the open-source software \gammapy. As highlighted by Sect.~\ref{sec:joint_crab}, a common data format and shared analysis tools allow multi-instrument joint analysis and effective data sharing. This synergy between experiments is particularly relevant given the complementary nature of pointing and wide-field instruments. This will be specially relevant for the joint scientific exploitation of future observatories such as SWGO and CTA.

\section{Discussion}
\label{sec:future}
The future of data formats in gamma-ray astronomy will very likely be linked to the future of the GADF initiative. As discussed over the text, this community-driven initiative has proposed the first available set of specifications for high-level data for the current and next generation of ground-based gamma-ray instruments. In this section we will discuss the main limitations affecting current specifications, as well as foreseeable ways in which they will evolve over the next decade.
\par
One of the main drivers of the evolution and improvement of the GADF will be the target requirements imposed by the future ground-based observatories. These will impose high-level data (and especially, the IRFs) to be described and parameterised in more complex ways, directly benefiting also the current generation of instruments. Possible extensions of the format to meet these requirements could include: a better field of view binning approach, removing the assumption of radial symmetry; inclusion of time dependency in the IRF components; distinguishing between different event types based on the hardware, reconstruction or analysis settings. Mature format specifications will be crucial for defining and testing current instruments legacy data, as they face the challenge of digesting decades of data (taken by instruments with evolving capabilities) and ensuring their proper use and interpretation. 
\par
In order to confront these challenges and to ensure the long-term feasibility of the GADF specifications, a more formal governance structure is needed. For this reason, a body of representatives from the high-energy ground-based community will be defined to act as a coordination committee. This governance definition effort, currently in progress, will inherit from the evolution of similar community-driven initiatives (for instance, the Astropy Project role responsibilities \cite{astropy_team}). 
\par
Even if the GADF specifications were inspired by high-energy satellites and primarily developed by and for the IACT community, they are able to represent high-level data products from other event-based high-energy astrophysical instruments. As shown in Sect.~\ref{subsec:hawc}, other high-energy gamma-ray observatories such as WCD (like HAWC or the future SWGO) naturally fit the GADF specifications, allowing the use of available open-source data analysis tools. In the coming years, the inclusion of other observatories will be explored, especially in the context of high-energy multi-messenger astronomy: allowing the inclusion of data from neutrino or even gravitational wave observatories would require some changes to the specifications, but at the same time would naturally allow the use of common science tools for joint multi-messenger analyses.

\section{Conclusions}
\label{sec:conclusions}

This review presented an outlook on the evolution of the data format in VHE gamma-ray astronomy from private and diverse specifications to the open and standardised ones proposed under the GADF initiative. The GADF initiative is presented as a community-driven effort to provide a common and open high-level data format for gamma-ray instruments. The specifications proposed within the GADF refer to high-level data products that would allow the production of scientific results: they are independent of the particular detection technique, thus allowing to accommodate data from different telescopes (e.g. IACT and WCD). The format definition was driven by the requirement to operate the next generation of gamma-ray instruments (such as CTA) as open observatories, with the consequent need of providing non-expert external users with open data products that are easy to interpret. Another aspect of this demand was the development of open-source gamma-ray data-analysis tools, whose evolution is now also linked to the data standardisation effort. 
\par
Current GADF specifications have proven to be robust by several publications analysing GADF-compliant data with these open-source science tools, validating their results against those obtained with the established closed-source software in use by current collaborations. These publications confirmed not only the correctness of the information incorporated in the format specifications but, at the same time, the capabilities of this new generation of open-source science tools. Other publications have instead proven the feasibility of multi-instrument and fully-reproducible analyses once the common format and open software are used. Even if future instruments are driving the open data and software development, the current generation can significantly benefit from their advancement. Their adoption ensures a larger user and maintainer base for the legacy data of current instruments, and eventually more sophisticated data storage and analysis techniques. The H.E.S.S. collaboration already pioneered a first public release of GADF-compliant data. All currently operating VHE gamma-ray experiments are nowadays also able to produce GADF-compliant data products, though for the moment they have mostly been used internally. Multi-instrument scientific projects using these data products are on their way, sharing data among collaborations through the use of memoranda of understanding. 
\par
The standardisation effort remains open to the inclusion not only of more gamma-ray instruments but also of telescopes observing the universe with other messengers. With the initiative being community-driven, high-energy astrophysicist in need of new extensions to the format are able to propose them. The recent efforts reviewed in this issue successfully employing GADF-compliant data and open-source analysis tools will surely foster their usage for further scientific projects. The GADF does not represent an isolated effort and aims at maintaining compatibility with other established standards in high-energy astronomy, like the OGIP (on which the GADF largely draws), or those used for high-level products within the Virtual Observatory \cite{ivoa}. Promoting the use of open-source analysis tools as well as common open data formats will distinguish high-energy astrophysics in the future as one of the few branches of modern science unconcerned by the reproducibility dilemma affecting many other disciplines \cite{repro_crisis}.

\vspace{6pt} 



\funding{This work was supported by the European Commission’s Horizon 2020 Program under grant agreement 824064 (ESCAPE - European Science Cluster of Astronomy \& Particle Physics ESFRI Research Infrastructures), by the the ERDF under the Spanish Ministerio de Ciencia e Innovaci{\'o}n (MICINN, grant PID2019-107847RB-C41), and from the CERCA program of the Generalitat de Catalunya.}

\acknowledgments{We are grateful to Gernot Maier, Maximilian N{\"o}the, Bruno Kh{\'e}lifi and Karl Kosack for their suggestions and for reviewing the manuscript.}

\conflictsofinterest{The authors declare no conflict of interest.} 





\end{paracol}
\reftitle{References}


\externalbibliography{yes}
\bibliography{biblio.bib}


%


\end{document}